\documentclass[12pt]{article}

\usepackage{float}
\usepackage[utf8]{inputenc}
\usepackage{cancel}
\usepackage[left=0.75in,right=0.75in,top=0.75in,bottom=0.75in]{geometry}
\usepackage{setspace}
\usepackage{graphics}
\usepackage{amssymb}
\usepackage{units}
\usepackage{amsmath}

\usepackage{bm}
\usepackage{mathtools}
\usepackage{times}
\usepackage{color}
\usepackage{cite}
\usepackage{graphicx}
\usepackage[normalem]{ulem}

\newcommand{\dee}{{\rm d}}
\newcommand{\calS}{{\mathcal S}}

\newcommand{\calD}{{\mathcal D}}
\newcommand{\calJ}{{J}}

\newcommand{\calP}{{\mathcal P}}
\newcommand{\eqa}{\begin{eqnarray}}
\newcommand{\eeqa}{\end{eqnarray}}
\newcommand{\eq}{\begin{equation}}
\newcommand{\eeq}{\end{equation}}

\newcommand{\di}[1]{\mathop{\mathrm{d}#1}}

\renewcommand{\Re}{{\rm Re}\,}
\renewcommand{\Im}{{\rm Im}\,}

\newcommand{\ito}{It\^o }
\newcommand{\drm}{\mathop{}\!\mathrm{d}}
\DeclareMathOperator{\sech}{sech}

\newcommand{\edit}[1]{\textcolor{black}{#1}}

\begin{document}

\title{Complex paths for real stochastic processes}

\author{D.A. Baldwin$^1$, A.J. McKane$^2$, and S.P. Fitzgerald$^{1}$\thanks{S.P.Fitzgerald@leeds.ac.uk}\\$^1$Department of Applied Mathematics, University of Leeds, \\LS2 1JT, United Kingdom\\
   $^2$Department of Physics and Astronomy, University of Manchester,\\ M13 9PL, United Kingdom}

\maketitle 

\begin{abstract}
The calculation of the decay rate of a metastable state in the path-integral formulation of stochastic processes is revisited. Previous derivations of this rate were achieved at the cost of a step that is difficult to justify mathematically. We show that this difficulty can be resolved by working with a \edit{complex} extremal solution that arises naturally in the It\^{o} formulation of the path integral. To make the analysis as transparent as possible, we choose a simple potential for which the extremal solution can be written in terms of elementary functions. The mechanism identified here, however, is not restricted to this example and holds more generally.
\vspace{5mm}

\end{abstract}

\begin{spacing}{1.25}

\section{Introduction}
The decay of a metastable state of a system is a ubiquitous process in the
sciences, which in many classical systems can be modelled by motion in a
potential subject to noise. In the simplest case, where the noise is white
and where the system has only one degree of motion, the rate of decay can be
calculated from the Fokker--Planck equation describing the time evolution of
the system \cite{risken,gardiner}, and is given by the classic formula of
Kramers \cite{kramers}. However, if the system has an infinite number of
degrees of freedom, for instance in the case of the decay of a field, or if
the noise is not white, then a simple Fokker--Planck equation may not exist,
and the description of the system in terms of path integrals is frequently
more useful \cite{graham,mckane_review,wio}.

In the case of a field, the formalism required to study the decay of a
metastable state was developed by Langer \cite{langer1967}. His central
object of study was a droplet of the stable state in an initially prepared
metastable state. This droplet corresponds to field configurations in the
path integral which were later called instantons in the context of quantum
field theory \cite{coleman}. Visualising decays of metastable states in this
way is also useful in systems with just a few degrees of freedom, and
instantons are useful both conceptually and calculationally in these
situations \cite{caroli,baibuz,mckane3}. \edit{Rigorous progress has also been made in calculating such decay rates in the context of stochastic partial differential equations \cite{berglund}.}

While the Kramers rate formula 
can be calculated using instantons \cite{mckane,mckane3,einchcomb1995}, there
is one step in the derivation which remains unsatisfactory. In the course of
the calculation one has to carry out an integration over the separation
between an instanton and an anti-instanton, which attract each other. This
leads to a divergent integral. Various ways of getting around this problem
have been put forward, most of them involving an analytic continuation from
$D$ to $-D$, where $D$ is the diffusion constant \cite{bogomolny1980,zinnjustin1981}. However, none of
these proposals has been entirely satisfactory. Here we give a resolution of
this difficulty, inspired by analogous computations carried out in a quantum
mechanical setting \cite{behtash2015b,behtash2015, behtash2016,behtash2018}. In the stochastic case, the solution turns out to arise naturally from the It\^{o}
prescription in the path integral, which leads to a modified variational path.

The outline of the paper is as follows. In Section 2, we briefly review the
formalism we use, and in Section 3 the extremal solution of interest and its
action are determined. We discuss the important differences arising from the choice of discretization for the underlying stochastic differential equation, i.e. \ito or Stratonovich. The contributions which give rise to the prefactor in
the Kramers rate formula are discussed in Section 4 and the features which
make the calculation well-defined are explained in Section 5. We conclude in
Section 6. Appendices outline various technical aspects
of the calculation. \edit{We aim to keep the presentation accessible to statistical physicists, which means that some rigour is sacrificed, so we have added references to the mathematical literature to address this.}

\section{Formalism}
We begin with the overdamped one-dimensional Langevin equation describing a particle in a potential $V(x)$, subject to additive Gaussian white noise of strength $D$:
\eq \label{eq:langevin}
\dot x(\tau) = -V'(x(\tau)) + \xi(\tau).
\eeq
Here, $V'(x)=\mathrm{d}V/\mathrm{d}x$ and $\dot x=\mathrm{d}x/\mathrm{d}\tau$ where $\dot{x}(\tau)$ is understood as the continuum limit of a discrete-time difference. We rescale time so that the friction coefficient is unity.
The probability density functional for the noise is given by
\eq
\calP[\xi] \propto \exp\left( -\frac1{4D}\int_0^t \xi(\tau)^2\,\dee\tau\right),
\eeq where the integral is interpreted as the limit of discrete time steps. Substituting $\xi$ from eq.~\eqref{eq:langevin} yields
\eq
\calP[x] \propto \exp\left( -\frac1{4D}\int_0^t \left(\dot x + V'(x) \right)^2\dee\tau\right),
\eeq and the integral in the exponent is the stochastic action
\eq
\calS[x] := \int_{0}^{t}\bigl(\dot x(\tau)+V'(x(\tau))\bigr)^2\,\dee\tau,
\label{eq:action}
\eeq
which is a functional of the path $x(\tau)$  \cite{onsager,stratonovich,graham,wio}. \edit{For more mathematically rigorous treatments, see e.g. \cite{lu,durr,takahashi,fujita}}.

The transition probability $P(X_t = x|X_0 = x_0)$ can be cast as an integral over system trajectories
\eq
P(x,t|x_0,0) = \int \calD x\,\calJ[x]\,\exp\left( -\frac{\calS[x]}{4D}\right),\label{eq:PI1}
\eeq taken over trajectories $x$ satisfying $x(0)=x_0;x(t) = x$ \cite{graham,wio}. \edit{The functional Jacobian $\calJ$, required when changing from the Wiener measure $\calD\xi$ to the path measure $\calD x$ in eq.~\eqref{eq:PI1}, depends on the discretization scheme used to define the path integral. When the drift $V'(x)$ in the Langevin equation~\eqref{eq:langevin} is evaluated at $x_n + \alpha(x_{n+1} - x_n)$ on the time-slice from $\tau_n$ to $\tau_{n+1}$, the Jacobian is
\begin{equation}
\calJ[x] = \exp\!\left( \alpha \int_{0}^{t} V''(x(\tau))\,\dee\tau \right),
\label{eq:jacobian}
\end{equation}
with $\alpha \in [0,1]$ \cite{wio}. The choices $\alpha = 0$ and $\alpha = 1/2$ correspond to the \ito and Stratonovich prescriptions respectively. In the \ito case the Jacobian is one, whereas in the Stratonovich case it is $\exp\!\big(\tfrac{1}{2}\int_0^t V''(x)\,\dee\tau\big)$.}

\edit{For additive white noise the Langevin eq.\eqref{eq:langevin} specifies the density of the process unambiguously, so the path integral eq.\eqref{eq:PI1} must be independent of $\alpha$. The mechanism by which this independence is realized, however, differs between conventions: the same $V''$ contribution that appears explicitly in the Stratonovich Jacobian emerges in the \ito formulation, via \ito's lemma applied to a term in the action, $\mathrm{d} V'(x(\tau))/\mathrm{d}\tau$, as we discuss in the following section. The two organizations of the calculation are equivalent, but they are not identical at the level of the saddle-point structure used in the weak noise approximation. The key difference is that in the \ito formulation, the $DV''$ contribution sits inside the action to be minimized, and hence affects the minimum action extremal paths, whereas in Stratonovich, the term is excluded from the action, and the extremal paths are independent of $D$. The excluded term is then calculated using these $D$-independent extrema according to eq.\eqref{eq:jacobian}.
The \ito organization is the one that admits an exact complex extremal trajectory, and is the focus of our paper. Recent experimental results \cite{gladrow} suggest that the $V''$ term influences the most probable paths themselves (and not just the probabilities associated with them), suggesting that the \ito approach developed below is the appropriate one. }

\subsection{Weak noise limit}\label{weak noise lim}

\edit{In this section, we show how the $V''$ term arises from the \ito integral in $\calS$.} As $D\to 0$, the dominant contributions to the functional integral in eq.\eqref{eq:PI1} come from paths near stationary points of the action functional  \cite{FW,ludwig1975}. In this regime we apply a steepest-descent approximation by expanding the action to quadratic order about a stationary path, $x_\star(\tau) $ say, which is symbolically written as:
\begin{equation}\label{eq:functional-taylors}
\mathcal{S}[x]
=
\mathcal{S}[x_\star+y]
=
\mathcal{S}[x_\star]
+\frac{1}{2}\frac{\delta^2\mathcal{S}}{\delta x^{2}}\bigg|_{x=x_\star}y^{2}
+\cdots,
\end{equation}
and performing the resulting Gaussian functional integral over $y(\tau)$. The stationary path $x_\star(\tau)$ is defined by the variational condition
\begin{equation}\label{eq:EL_general}
\frac{\delta\mathcal{S}}{\delta x}\bigg|_{x=x_\star}=0,
\end{equation}
which yields the Euler--Lagrange equation.

The Euler--Lagrange equation depends on the discretization scheme used to define the functional integral \cite{graham,cugliandolo2019}. In particular, the It\^{o} scheme produces an additional $V''$ term, which modifies the effective potential term in the action. In the cubic case we consider here, this induces a small \emph{tilt} in the effective potential and, as we show below, it admits new trajectories.
To see how this tilt enters, consider the action functional
\begin{equation}\label{eq:OM-action}
\mathcal{S}[x]=\int_{0}^{t}L\big(x(\tau),\dot{x}(\tau)\big)\,\mathrm{d}\tau,
\qquad
L\big(x,\dot{x}\big):=\big(\dot{x}+V'(x)\big)^{2},
\end{equation}
with Dirichlet boundary conditions $x(0)=x_{0}$ and $x(t)=x_{1}$. Since $L$ has no explicit dependence on $\tau$, the associated Hamiltonian
\eq
H := p\,\dot x - L, \qquad p := \frac{\partial L}{\partial \dot x},
\eeq
is conserved along any stationary path (i.e., along solutions of the Euler--Lagrange equation), where $p$ is the conjugate momentum. For the Lagrangian eq.\eqref{eq:OM-action} one finds
\eq
p = 2\big(\dot x + V'(x)\big),
\qquad
H = \dot x^{\,2} - V'(x)^{2}.
\label{eq:H-def}
\eeq
We refer to $H$ as the (conserved) energy of the corresponding effective-mechanics problem. Let $x$ be a stationary path of $\mathcal{S}$, and take a test variation $x_{\varepsilon}=x+\varepsilon\eta$, where $\eta(0)=\eta(t)=0$ and $|\varepsilon|\ll1$. The first variation is
\begin{equation}\label{eq:first-var-S}
\delta\mathcal{S}
=2\int_{0}^{t}
\big(\dot{x}+V'(x)\big)\big(\dot{\eta}+V''(x)\eta\big)\,\mathrm{d}\tau.
\end{equation}
Integrating the $\dot{\eta}$ term by parts, with respect to the smooth test function $\eta$ and vanishing boundary term due to $\eta(0)=\eta(t)=0$, yields
\begin{equation}\label{eq:IBP-eta}
\int_{0}^{t}\big(\dot{x}+V'(x)\big)\dot{\eta}\,\mathrm{d}\tau
=-\int_{0}^{t}\eta\,\frac{\mathrm{d}}{\mathrm{d}\tau}\big(\dot{x}+V'(x)\big)\,\mathrm{d}\tau.
\end{equation}
The discretization convention enters through the evaluation of $\frac{\mathrm{d}}{\mathrm{d}\tau}V'(x(\tau))$ along the stochastic path $x(\tau)$. By It\^{o}'s lemma (see e.g. \cite{oksendal2003,karatzas1991}), 
\begin{equation}\label{eq:ito-chain}
\frac{\mathrm{d}}{\mathrm{d}\tau}V'(x(\tau))
=V''(x)\,\dot{x} + D\,V'''(x),
\end{equation}
so that substituting eq.\eqref{eq:ito-chain} into eq.\eqref{eq:first-var-S}--eq.\eqref{eq:IBP-eta} and noting that the $V''(x)\dot{x}$ terms cancel, we obtain
\begin{equation}\label{eq:deltaS-final}
\delta\mathcal{S}
=2\int_{0}^{t}\eta(\tau)\,\Big(-\ddot{x}+V'(x)V''(x)-D\,V'''(x)\Big)\,\mathrm{d}\tau.
\end{equation}
Since $\eta$ is arbitrary, the stationarity condition $\delta\mathcal{S}=0$ leads to the It\^{o} Euler--Lagrange equation
\begin{equation}\label{eq:Ito-EL}
\ddot{x}(\tau)=V'(x(\tau))\,V''(x(\tau)) - D\,V'''(x(\tau)).
\end{equation}
By contrast, in the Stratonovich convention, the standard chain rule applies directly, $\frac{\mathrm{d}}{\mathrm{d}\tau}V'(x)=V''(x)\dot{x}$, yielding simply $\ddot{x}(\tau)=V'(x(\tau))V''(x(\tau))$ with no $D$-dependent correction. The Itô Euler--Lagrange equation eq.\eqref{eq:Ito-EL} can be expressed in terms of an effective potential, 
\begin{equation}\label{eq:eff-potential}
\ddot{x}(\tau) = - \frac{1}{2} U'(x(\tau)), \quad U(x) := - V'(x)^2 + 2DV''(x),
\end{equation}
which is the Newtonian equation of motion for a particle of mass 2 moving in the potential $U$. The $DV''$ term modifies the effective Hamiltonian mechanics, altering the extremal trajectories relative to the Stratonovich case, where the potential is $U = -V'^2$.

We are minimizing the same functional eq.\eqref{eq:action}, but the saddle-point approximation depends on the discretization scheme employed. As is well known for diffusive processes driven by additive white noise, either choice gives the same result if the path integral is evaluated exactly \cite{graham,cugliandolo2019}. However, the two schemes differ in how they organize the approximation. In particular, the extremal trajectories differ through the $D$-dependent term in the It\^{o} variational equations. The Stratonovich scheme yields zero-energy extremals in the infinite-time limit $t\to\infty$, which are called instantons. These satisfy the first-order equation $\dot{x}=\pm V'(x)$, whereas no such reduction to a first-order ODE is available in the It\^{o} convention. Consequently, the perturbative contributions around the extremal are reordered between the two schemes. \edit{Note that the $V''$ term is not neglected in either scheme. In the approach above, it emerges naturally as a consequence of the \ito discretization. In the Stratonovich scheme, though it does not affect the extremals (instantons), it is calculated after the action minimization step by substituting the extremal in \eqref{eq:jacobian}}. 

\subsection{Choice of model}
To keep the analysis analytically tractable, we begin with a non-confining cubic potential, for which the effective potential $U$ given in eq.\eqref{eq:eff-potential} is a quartic polynomial and the relevant extremal trajectories can be obtained in closed form. The cubic potential is the archetypal model for a stochastic barrier-crossing process, and its non-confining nature with no equilibrium density brings the difficulties mentioned in the introduction to the fore. By fortunate coincidence, the quartic effective potential $U$ that results is almost identical to the inverted bosonic potential considered in the supersymmetric quantum mechanical context of Refs.\cite{behtash2015b,behtash2015,behtash2016,behtash2018}, and the complexified extremal paths and their actions can be computed in precisely the same way, \edit{including the careful treatment of fluctuations around the extremal path. Whilst these calculations are similar, the physical process under consideration, the origin of the modification of the effective potential, and our main result, are completely independent. We include the full detail of the calculations to keep the presentation self-contained and accessible to the intended readership.} The same framework extends, in principle, to more complicated potentials and to higher dimensions.
\begin{figure}
\centering
\includegraphics[width=0.8\textwidth]{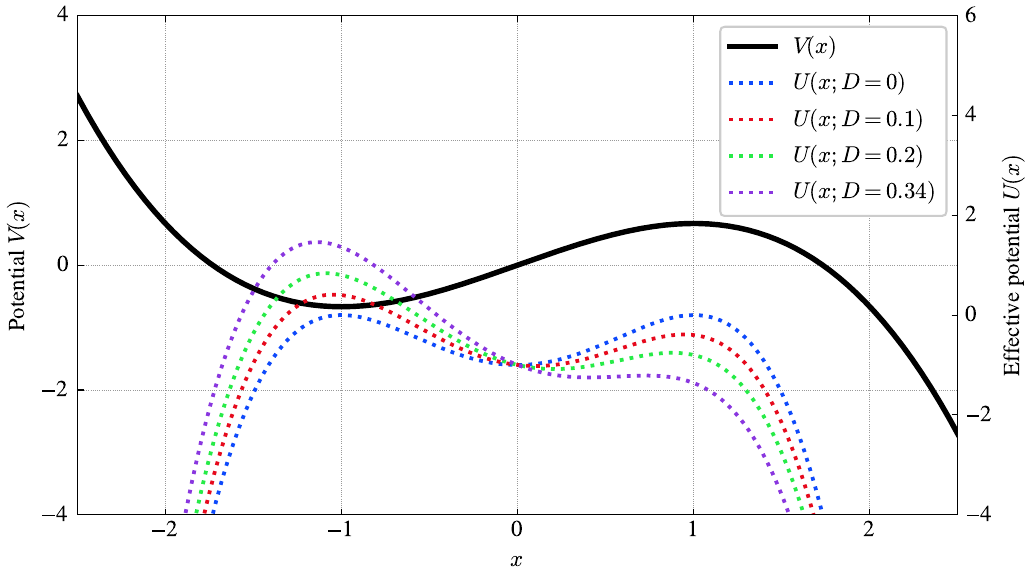}
\caption{Potential $V$ and effective potential $U$ for four values of the noise strength $D$. Note that $U$ and $V$ are plotted on different $y$ axes to emphasize the shape of $U$.}
\label{fig:UV}
\end{figure}
We consider the potential
\eq
V(x) = -\frac{x^3}{3} + x + \frac{2}{3}
\eeq 
as a model for a barrier crossing process. The potential $V$ has only one metastable minimum at $x=-1$ and a barrier of height $ E_b = \Delta V = V(1) - V(-1) = \frac{4}{3}$ at $x=1$. The potential is shifted so that $V$ is zero at the minimum. The effective potential is given by 
\eq
U(x) = -(x^2-1)^2 - 4Dx,
\eeq 
and this is where the Hamiltonian dynamics take place. Fig.\ref{fig:UV} shows $V$ and four variants of $U$ with $D=0,0.1,0.2$ and $0.34$. Having identified the classical mechanics problem in the tilted potential $U$, we can now study the relevant extremal trajectories and develop the weak-noise expansion of the stochastic path integral.
\section{Determination of the extremal solution and its action}
The first step is to find the least-action path $x_{\rm cl}$ which will satisfy eq.\eqref{eq:Ito-EL} and then evaluate the action on this trajectory. This controls the leading contribution in the weak-noise limit,
\eq \label{eq:leading-order-cont}
\calP \sim \exp\!\left(-\frac{\mathcal{S}_{\rm cl}}{4D}\right),
\eeq 
where $\mathcal{S}_{\rm cl}:=\mathcal{S}[x_{\rm cl}]$ denotes the action evaluated on the path $x_{\rm cl}$. We refer to $x_{\rm cl}$ and $\mathcal{S}_{\rm cl}$ as the classical solution and classical action, respectively. We first summarize the Stratonovich picture \edit{as covered for example in} \cite{mckane3} and then contrast it with the It\^{o} formulation. Prefactors from fluctuations and the Jacobian are considered in Section 4.
\subsection{Stratonovich effective potential and its drawback}
First, consider the Stratonovich discretization. In this formulation, the effective potential takes the $D$-independent form,
\eq
U(x)=-V'(x)^2=-(x^2-1)^2,
\eeq
and the zero-energy extremal trajectories satisfy the first-order equations
\eq
\dot{x}=\pm V'(x)=\pm(1-x^2).
\eeq
We refer to the right-moving zero-energy trajectory connecting $x=-1$ to $x=+1$ as the instanton, and to its time-reversed partner connecting $x=+1$ to $x=-1$ as the anti-instanton. In the present cubic model these are
\eq
x_I(\tau)=\tanh\tau,
\qquad
x_A(\tau)=x_I(-\tau)=-\tanh\tau,
\eeq
where the subscripts $I$ and $A$ denote instanton and anti-instanton, respectively. The extremal trajectory relevant to the return transition, where the particle starts at $x_0$ at time $0$ and returns to $x_0$ at time $t$, can then be approximated for large $t$ by a widely separated instanton-anti-instanton pair,
\eq \label{eq:IA-approx}
x(\tau)\approx -1 + \tanh(\tau-t_1)-\tanh(\tau-t_2),
\qquad t_2>t_1.
\eeq
The order of the two events is a right-moving instanton followed by a left-moving anti-instanton.

This step exposes a basic difficulty: the instanton and anti-instanton attract, so the action decreases as their separation shrinks, and the integral over the separation diverges in the $t\to\infty$ limit. One way to render this integral meaningful is to change the sign of $D\to -D$, which makes the interaction repulsive and the separation integral convergent (see, e.g., \cite{bogomolny1980,zinnjustin1981,mckane3}). While this uncontrolled “analytic continuation” in $D$ turns out to identify the correct steepest-descent contour, the sign change is mathematically unsatisfactory. In what follows, we avoid this by fixing the integration contour geometrically after complexifying the path space \cite{behtash2015}.

There is also a conceptual drawback of the Stratonovich effective potential $U=-V'(x)^2$: it does not distinguish minima from maxima of $V$. In particular, $V$ and $-V$ generate the same $U$, and every critical point of $V$ is mapped to a zero of $U$. Consequently, the effective-mechanics particle can spend comparable time near the barrier top and near the well bottom, which is not physically satisfactory in the present non-confining barrier-crossing setting. This issue is resolved once the It\^{o} correction is retained in $U$.

\subsection{\ito\ effective potential and extremal structure} 

Now consider the effective potential $U$ in the It\^{o} convention. The additional $2D\,V''$ term raises $U$ near minima of $V$ and lowers it near maxima, and it shifts the extrema of $U$ slightly away from the critical points of $V$ (see Fig.~\ref{fig:UV}). As a consequence, the zero-energy trajectories that underpin the Stratonovich instanton picture no longer support infinite-time motion. In the tilted potential, a trajectory can spend infinite time only by approaching a stationary point of $U$ with zero velocity. Thus, any infinite-time solution must approach a local maximum of $U$ asymptotically, and its conserved energy differs from zero by an amount of order $D$.

For $D$ below the critical value at which $U$ ceases to have two local maxima ($D_c=\sqrt{4/27}\approx0.38$), there is a real ``bounce'' trajectory that starts at rest at the right-hand maximum of $U$ (near $x=1$), moves left, spends a long but finite time near the well region around $x=-1$, and then returns to its starting point. However, this is not the trajectory relevant to the return transition based at $x_0\approx-1$: if one attempts to splice this motion into a path that starts and ends near $x=-1$, the resulting trajectory necessarily spends infinite time near the barrier top and only finite time near the well bottom. We instead require a return trajectory that starts from rest near the left-hand maximum of $U$ (near $x=-1$). For real trajectories, this is impossible: once such a solution moves to the right, it crosses the barrier region and accelerates toward $x\to+\infty$ without turning back. Therefore, to realize a return path satisfying the required boundary conditions, we must allow the extremal trajectory to leave the real line and seek solutions in complexified path space. Accordingly, we complexify the coordinate $x\mapsto z$ and consider trajectories $z(\tau)\in\mathbb{C}$. \edit{The complexification of the path was partially anticipated in the stochastic context by \cite{caroli}, and in the quantum case in chapter 17 of \cite{kleinert}, though neither found the exact complex solution. Reference \cite{caroli} identified the correct contour heuristically, noting that a rigorous derivation was precluded by the ``absence of a general theory of analytic continuation for functional integrals''. The exact solution and the associated analytic continuation were obtained, for the quantum case only, in refs. \cite{behtash2015b,behtash2015,behtash2016,behtash2018}.}

To understand how such return solutions can arise, we examine the turning-point condition more carefully. In the effective-mechanics description, turning points are defined by the vanishing of the particle’s speed, i.e.\ $\dot z(\tau)=0$.
Using $H=\dot z^{\,2}+U(z)$, this is equivalent to the condition
\eq
\Phi(z):=H-U(z)=0,
\label{eq:TPE}
\eeq
where $H$ is the conserved energy discussed in eq.\eqref{eq:H-def}. For the cubic model, $U(z)=-(z^2-1)^2-4Dz$, so $\Phi$ is a quartic polynomial in $z$.

Now let $z_{\rm cr}$ denote the left-hand local maximum of $U$ (located near the real point $z=-1$). If we choose $H$ to equal the value of $U$ at this maximum, $H=U(z_{\rm cr})$, then $\Phi(z_{\rm cr})=0$. Moreover, since $z_{\rm cr}$ is a stationary point of $U$ we have $U'(z_{\rm cr})=0$, and hence $\Phi'(z_{\rm cr})=-U'(z_{\rm cr})=0$. Therefore, $z_{\rm cr}$ is a \emph{double root} of the quartic equation $\Phi(z)=0$.

The remaining two roots must then form either a second real double root or a complex conjugate pair. We choose the branch relevant to a return trajectory based at $z_{\rm{cr}}$: the remaining roots are a complex conjugate pair located near the barrier region $z\approx 1$. Consequently, there exists a pair of trajectories that start from rest at $z_{\rm cr}$, leave the real axis, turn at one of these complex turning points, and return to $z_{\rm cr}$. We call this complex solution $z(\tau)$ the ``complex bounce''.\footnote{This is referred to as a ``bion'' in the quantum mechanical context of \cite{behtash2016}.} (We refer to the corresponding trajectory starting and ending at the right-hand maximum near $z\approx 1$ as the ``real bounce'' as in \cite{behtash2016}. It again has a double root at the starting point, but $H<0$ and the two further roots are real, either side of the maximum near $z\approx -1$.)

\subsubsection{Complex bounce: closed-form solution and classical action}

We now calculate the complex bounce $z_{\rm cl}(\tau)$ and evaluate its classical action. Whilst the calculation of these quantities proceeds in the same way as given in the quantum mechanical context of ref. \cite{behtash2015b}, we include a detailed presentation here for completeness.  
From eq.\eqref{eq:TPE} and the preceding discussion, denoting the remaining turning points by $z_{\rm tu}^{\pm}$, we may factorize $\Phi$ as follows:  
\eq
\Phi(z)=(z-z_{\rm cr})^{2}(z-z_{\rm tu}^{+})(z-z_{\rm tu}^{-}).
\eeq
The complex bounce is the return trajectory that starts and ends at rest at $z_{\rm cr}$ and turns at
one of $z_{\rm tu}^{\pm}$. Fixing the time origin at the turning point, we impose
\eq \label{eq:cb-bcs}
z_{\rm cl}(\pm\infty)=z_{\rm cr},\qquad z_{\rm cl}(0)=z_{\rm tu}^{\pm}.
\eeq
Separating variables in $\dot z_{\rm cl}^{\,2}=\Phi(z_{\rm cl})$ yields the integral
\eq \label{eq:cb-quadrature}
\tau
= \pm\!\int_{z_{\rm tu}^{\pm}}^{z}\frac{\di{z'}}{\sqrt{(z'-z_{\rm cr})^{2}(z'-z_{\rm tu}^{+})(z'-z_{\rm tu}^{-})}}
=: F(z),
\eeq
subject to the boundary conditions in eq.\eqref{eq:cb-bcs}, where the branch of the square root is
chosen continuously along the trajectory. At this stage the solution is implicit, $\tau=F(z)$, and
inversion gives $z_{\rm cl}(\tau)=F^{-1}(\tau)$. The algebraic steps are given in \ref{App:CBmaterial}.
The final closed form and its action are
\begin{align}
z_{\rm cl}(\tau) &= z_{\text{cr}} + X(D)\!\left[\tanh\!\left(\frac{\omega}{2}(\tau + t_0)\right)
- \tanh\!\left(\frac{\omega}{2}(\tau - t_0)\right)\right],\label{eq:complex bounce}\\
\mathcal{S}_{\rm cl} &= \mathcal{S}[z_{\rm cl}]
= -H t + Y(D) + 16D \log\!\left(\sqrt{Z(D)}+\sqrt{Z(D)+1}\right),\label{eq:complex-action}
\end{align}
where $H$ is the conserved energy defined in eq.\eqref{eq:H-def}, and the frequency $\omega$, the
separation $t_0$, and the functions $X(D)$, $Y(D)$, and $Z(D)$ are defined in \ref{App:CBmaterial}. 
\begin{figure}
\centering
\includegraphics[width=0.8\textwidth]{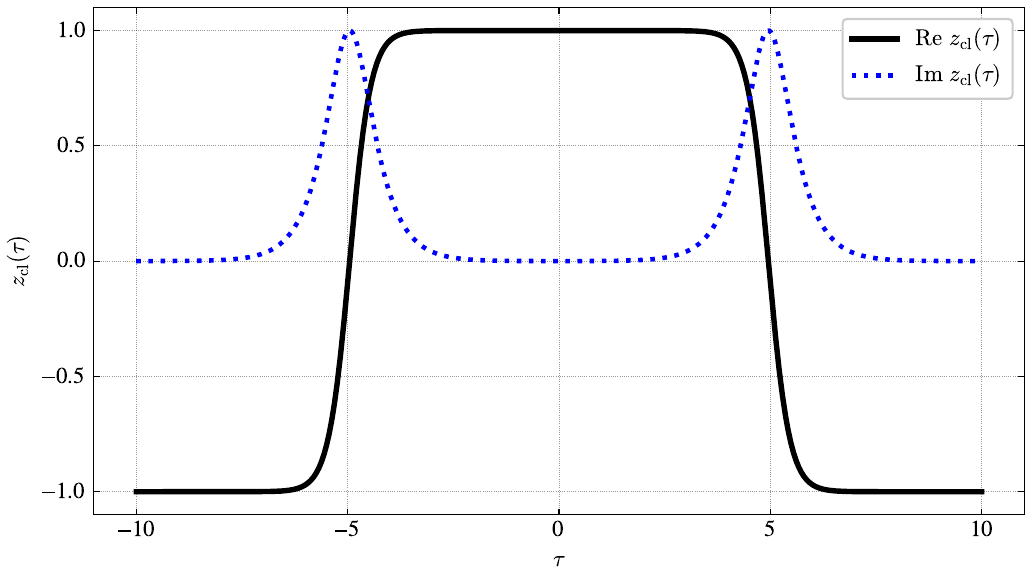}
\caption{Real (black solid) and imaginary (blue dotted) parts of the complex bounce eq.\eqref{eq:complex bounce} (the complex turning point is chosen to have positive imaginary part).}
\label{fig:complex-bounce}
\end{figure}

We emphasize that $z_{\rm{cl}}(\tau)$ is a genuine extremal of the action in \emph{complexified} path space: it satisfies the Euler--Lagrange equation eq.\eqref{eq:Ito-EL} with $x$ analytically continued to $z\in\mathbb C$. This step of complexification is explained in greater detail in Section 5. The critical point is given by
\begin{equation}\label{eq:z-critical}
z_{\text{cr}}= - \frac{2}{\sqrt{3}} \cos \left(\frac{1}{3}\arccos\left(\frac{3\sqrt{3}D}{2} \right) \right) = -1 - \frac{D}{2} + \frac{3D^2}{8} + \cdots,
\end{equation}
and the complex conjugate turning points are
\begin{equation}\label{eq:turning-points}
z_\text{tu}^{\pm} = - z_{\text{cr}} \pm i \sqrt{2D/|z_{\text{cr}}|} = 1 \pm i \sqrt{2D} + \frac{D}{2} + \cdots.
\end{equation}
The solution in eq.\eqref{eq:complex bounce} exhibits the complex bounce as an \emph{exact} instanton-anti-instanton configuration and the two $\tanh$ terms describe a pair with separation parameter $t_{0}$, which we define by
\begin{equation}\label{eq:t0-exact}
t_0 = \frac{2}{\omega} \operatorname{arcosh}\left(\sqrt{Q(D)+ \frac{1}{2}} \right).
\end{equation}
Crucially, this separation is \emph{complex-valued}. Taking the leading asymptotic form as $D\to 0$ reveals
\begin{equation}\label{eq:t0-asymptotic}
t_0 = \frac{1}{4} \left[\log \left(\frac{8}{D}\right) + (2 \sigma - 1) i \pi  \right] +\mathcal{O}(D),
\end{equation}
where $\sigma =$ 0 or 1 labels the two branches of the inverse hyperbolic cosine. The interpretation of this two-valuedness will be discussed in Section 5 in the context of quasi-zero mode integration. 

This solution $z(\tau)$ is an exact instanton-anti-instanton pair extremizing the complexified Hamiltonian dynamics in the effective potential $U(x) = -V'(x)^2 + 2DV''(x)$. By contrast, the composite instanton-anti-instanton path used in the Stratonovich approach is only an approximate solution to the real Hamiltonian mechanics in the simpler potential $U(x) =-V'(x)^2$, becoming exact only in the limit of infinite separation.

% --- continue directly after eq.\eqref{eq:complex-action} ---
Using the complex bounce action eq.\eqref{eq:complex-action}, the small-$D$ expansion of the classical action takes the form
\begin{equation}\label{eq:cb-action-expansion}
\mathcal{S}_{\rm cl}
=
-Ht+\frac{16}{3}
+4D\Bigl(1-\log\!\Bigl(\frac{D}{8}\Bigr)+ i(2\sigma-1)\pi\Bigr)
+\cdots.
\end{equation}
The leading contribution as $D\to 0$ is controlled by the constant term $16/3$. The logarithmic correction is of order $D\log D$ and therefore generates an algebraic factor in $\mathcal{P}$ via the exponential $\exp(-\mathcal{S}_{\rm cl}/4D)$ in eq.\eqref{eq:leading-order-cont}; it must be retained when computing the prefactor in section 4. The imaginary term in eq.\eqref{eq:cb-action-expansion} reflects the multivaluedness of the logarithm and will be discussed in section 5.

The term proportional to $Ht$ cancels against the normalization factor in the definition of $\mathcal{P}$ (see section 4). The dominant exponential behavior is therefore determined by the constant term $16/3$, giving
\begin{equation}\label{eq:leading_arrhenius}
\mathcal{P}\sim \exp\!\Bigl(-\frac{4}{3D}\Bigr)=\exp\!\Bigl(-\frac{\Delta V}{D}\Bigr),
\end{equation}
where $\Delta V=4/3$ (defined earlier), as expected.

By contrast, in the Stratonovich formulation the corresponding solutions satisfy $H\equiv 0$, and the classical action reduces to the constant $16/3$ without a $D\log D$ correction. We now turn to the next-to-leading order contribution, the prefactor.
\section{Fluctuations, normalization and the prefactor}
\label{sec:prefactor}

To obtain the prefactor we must account for quadratic fluctuations about the extremal trajectory. In the path-integral formulation it is convenient to divide by an exactly solvable harmonic reference problem, which removes the ill-defined overall normalisation constant. Let $x_{\rm a}$ denote the local minimum of the potential $V$, and consider the harmonic reference potential
\begin{equation}\label{eq:V0_def}
V_0(x)=\frac{1}{2}\,V''(x_{\rm a})\,(x-x_{\rm a})^{2},
\qquad
V''(x_{\rm a})>0.
\end{equation}
The corresponding conditional probability density $\mathcal{P}_0$ is given in eq.\eqref{eq:P0_OU}. We therefore consider the ratio
\begin{equation}\label{eq:P_normalised_def}
\frac{\mathcal{P}(x_1,t\mid x_0,0)}{\mathcal{P}_0(x_1,t\mid x_0,0)}
=
\frac{\displaystyle \int_{x(0)=x_0}^{x(t)=x_1}\!\!\mathcal{D}x\;
J[x]\exp\!\Big(-\frac{1}{4D}\,\mathcal{S}[x]\Big)}
{\displaystyle \int_{x(0)=x_0}^{x(t)=x_1}\!\!\mathcal{D}x\;
J_0[x]\exp\!\Big(-\frac{1}{4D}\,\mathcal{S}_{0}[x]\Big)} .
\end{equation}
In the It\^{o} formulation used here the Jacobian is one for all paths, whereas in Stratonovich formulation one has a nontrivial Jacobian factor $\exp\!\big(\tfrac12\int_{0}^{t}V''(x(\tau))\,\drm\tau\big)$. We therefore set $J=J_0=1$ from here on.

For the harmonic reference, the transition density reads
\begin{multline}\label{eq:P0_OU}
\mathcal{P}_0(x_1,t\mid x_0,0)
=
\sqrt{\frac{V''(x_{\rm a})}{2\pi D\,(1-\exp(-2V''(x_{\rm a})t))}}
\\[2pt]
\times
\exp\!\left[
-\frac{V''(x_{\rm a})}{2D\,(1-\exp(-2V''(x_{\rm a})t))}
\Big(x_1-x_{\rm a}-(x_0-x_{\rm a})\exp(-V''(x_{\rm a})t)\Big)^2
\right].
\end{multline}
We normalise by choosing equal endpoints at the minimum, $x_0=x_1=x_{\rm a}$ so the exponential factor in eq.\eqref{eq:P0_OU} reduces to unity. Hence,
\begin{equation}\label{eq:P0_def}
\mathcal{P}_0(t):=\mathcal{P}_0(x_{\rm a},t\mid x_{\rm a},0)
=
\sqrt{\frac{V''(x_{\rm a})}{2\pi D\,(1-\exp(-2V''(x_{\rm a})t))}}\,.
\end{equation}
To evaluate the ratio in eq.\eqref{eq:P_normalised_def} we expand the action to quadratic order about a contributing solution $z_{\rm cl}(\tau)$. Writing
\begin{equation}\label{eq:z_expand}
z(\tau)=z_{\rm cl}(\tau)+y(\tau),
\qquad
y(0)=y(t)=0,
\end{equation}
we obtain
\begin{equation}\label{eq:S_quadratic}
\mathcal{S}[z_{\rm cl}+y]
=
\mathcal{S}_{\rm cl}
+\langle y,\hat{\mathcal{M}}\,y\rangle
+\mathcal{O}(y^{3}),
\end{equation}
where $\hat{\mathcal{M}}$ is the fluctuation operator (subject to Dirichlet boundary conditions). The pairing is $\langle u\mid v\rangle:=\int_{0}^{t}u(\tau)\,v(\tau)\,\drm\tau$, so $\langle \dot z_{\rm cl}\mid \dot z_{\rm cl}\rangle=\int_{0}^{t}\dot z_{\rm cl}(\tau)^{2}\,\drm\tau$. In the $t\to\infty$ limit, time translation symmetry generates an exact zero mode (EZM) of $\hat{\mathcal{M}}$, namely $y_{\rm ezm}(\tau)\propto \dot z_{\rm cl}(\tau)$. We therefore introduce a collective coordinate $t_c$ \cite{zittartz1966,langer1967,gervais1975,mckane_wallace1978} (the complex bounce centre) and split the fluctuations as
\begin{equation}\label{eq:mode_split_ezm}
y(\tau)=a_{\rm ezm}\,y_{\rm ezm}(\tau)+y_\perp(\tau),
\qquad
y_{\rm ezm}(\tau)\propto \dot z_{\rm cl}(\tau),
\end{equation}
with $y_\perp$ in the complement of the EZM and $a_n$ the mode coefficients. The change of variables $a_{\rm ezm}\mapsto t_c$ produces a Jacobian factor
$\sqrt{\langle \dot z_{\rm cl}\mid \dot z_{\rm cl}\rangle/(4\pi D)}\int_{0}^{t}\drm t_c$.

On the complementary subspace we expand
$y_\perp(\tau)=\sum_{n\ge 1}a_n y_n(\tau)$ with $\hat{\mathcal{M}}y_n=\lambda_n y_n$ and perform the Gaussian functional integral. Each nonzero mode contributes a factor $(4\pi D/\lambda_n)^{1/2}$, so the result is expressed in terms of a functional determinant with the zero mode removed, denoted $\det'\hat{\mathcal{M}}$. The divergent overall constant is universal and cancels in the ratio eq.\eqref{eq:P_normalised_def} once the same procedure is applied to the quadratic reference problem, leaving a finite determinant ratio $\det'\hat{\mathcal{M}}/\det\hat{\mathcal{M}}_0$.

Collecting these pieces, and using the solution $\mathcal{P}_0(t)$ defined in eq.\eqref{eq:P0_def}, we obtain
\begin{equation}\label{eq:ito-path-integral_prefactor}
\mathcal{P}(x_1,t\mid x_0,0)
\sim
\mathcal{P}_0(t)\,
\exp\!\left(+\frac{\mathcal{S}_{0}}{4D}\right)
\exp\!\left(-\frac{\mathcal{S}_{\rm cl}}{4D}\right)
\sqrt{\frac{\langle \dot z_{\rm cl}\mid \dot z_{\rm cl}\rangle}{4\pi D}}
\int_{0}^{t}\!\drm t_c\;
\left(\frac{\det{}'\hat{\mathcal{M}}}{\det{}\hat{\mathcal{M}}_0}\right)^{-1/2},
\end{equation}
which becomes exact as $t \to \infty$. \edit{This is the \ito calculus analogue of the equivalent expression (eq.(30)) in ref.\cite{mckane3}, but each term is different.} Here $\mathcal{S}_0$ denotes the reference action evaluated on its extremal $x(\tau)=x_{\rm a}$, so that $\mathcal{S}_0=-2D\,V''(x_{\rm a})\,t$ (the remaining terms vanish since $\dot x=0$ and $V_0'(x_{\rm a})=0$). A key piece in eq.\eqref{eq:ito-path-integral_prefactor} is the functional determinant with the EZM removed, which we now look at in more detail. 

\subsection{Fluctuation operator for the complex bounce}
\label{subsec:fluct_operator}

We now write the fluctuation operator $\hat{\mathcal{M}}$ appearing in the quadratic expansion in eq.\eqref{eq:S_quadratic} explicitly. In the notation of the functional Taylor expansion in eq.\eqref{eq:functional-taylors}, the second variation defines the integral kernel
\begin{equation}\label{eq:M_kernel_def}
\left.\frac{\delta^2\mathcal{S}}{\delta z(\tau)\,\delta z(\tau')}\right|_{z=z_{\rm cl}}
=:\hat{\mathcal{M}}(\tau,\tau')
=\hat{\mathcal{M}}_{\tau}\,\delta(\tau-\tau'),
\end{equation}
so that, for $z(\tau)=z_{\rm cl}(\tau)+y(\tau)$ with $y(0)=y(t)=0$,
\begin{equation}\label{eq:S_quad_op}
\mathcal{S}[z_{\rm cl}+y]
=
\mathcal{S}_{\rm cl}
+\int_{0}^{t}\!\drm\tau\; y(\tau)\,
\Bigl[-\frac{\drm^{2}}{\drm\tau^{2}}-\frac{1}{2}\,U''\!\bigl(z_{\rm cl}(\tau)\bigr)\Bigr]\,
y(\tau)
+\mathcal{O}(y^{3}),
\end{equation}
with $U(z)=-V'(z)^{2}+2D\,V''(z)$. Hence
\begin{equation}\label{eq:M_def}
\hat{\mathcal{M}}
=
-\frac{\drm^{2}}{\drm\tau^{2}}-\frac{1}{2}\,U''\!\bigl(z_{\rm cl}(\tau)\bigr)
=
-\frac{\drm^{2}}{\drm\tau^{2}}
+V''(z_{\rm cl})^{2}+V'(z_{\rm cl})V'''(z_{\rm cl})-D\,V''''(z_{\rm cl}),
\end{equation}
and we drop the subscript $\tau$ on the differential operator.

For the cubic potential, substituting the complex bounce solution from eq.\eqref{eq:complex bounce} gives the explicit Schr\"odinger form
\begin{equation}\label{eq:fluct-operator}
\hat{\mathcal{M}}
=
-\frac{\drm^2}{\drm\tau^2}
+\omega^2
\left[
1-\frac{3}{2}
\left(
\sech^2\!\left(\frac{\omega}{2}(\tau+t_0)\right)
+
\sech^2\!\left(\frac{\omega}{2}(\tau-t_0)\right)
\right)
\right].
\end{equation}
This has the form of two well-separated wells centered at $\tau=\pm \Re t_0$. A short derivation of eq.\eqref{eq:fluct-operator} is given in~\ref{App:fluct_operator_derivation}.

The determinant ratio in eq.\eqref{eq:ito-path-integral_prefactor}, with the exact zero mode removed, can be evaluated by one-dimensional determinant formulas in the real bounce problem, where the fluctuation operator is Hermitian \cite{tarlie}. Let $\hat{\mathcal{M}}_{\rm rb}$ denote the fluctuation operator about the real bounce. Using the formula and notation summarized in~\ref{App:RBdet}, we obtain
\begin{equation}\label{eq:det-ratio-rb-closed}
\lim_{t\to\infty}\frac{\det'\hat{\mathcal{M}}_{\rm rb}}{\det\hat{\mathcal{M}}_0}
=
-
\frac{\langle \dot{x}_{\rm rb}\mid \dot{x}_{\rm rb}\rangle}
{32 X_{\rm rb}^{2}\omega_{\rm rb}^{3}\sinh^{2}(\omega_{\rm rb} t_{0,{\rm rb}})},
\end{equation}
and in the weak-noise limit,
\begin{equation}\label{eq:det-expansion-rb}
\lim_{t \to \infty}\frac{\det'\hat{\mathcal{M}}_{\rm rb}}{\det\hat{\mathcal{M}}_0}
=
-\frac{D}{512}+\mathcal{O}(D^2).
\end{equation}

For the complex bounce, however, the fluctuation operator $\hat{\mathcal{M}}$ is complex symmetric rather than Hermitian, so these formulas do not apply directly. We therefore define the complex bounce determinant ratio by analytic continuation of the corresponding real bounce expression under $D\mapsto -D$\footnote{Note that this is an entirely innocuous analytical continuation that does not affect the convergence of any integral.}:
\begin{equation}\label{eq:det-ratio-cb-def}
\lim_{t \to \infty}\frac{\det'\hat{\mathcal{M}}}{\det\hat{\mathcal{M}}_0}
\;:=\;
\left.
\left(
\lim_{t \to \infty}\frac{\det'\hat{\mathcal{M}}_{\rm rb}}{\det\hat{\mathcal{M}}_0}
\right)
\right|_{D\mapsto -D}.
\end{equation}
Under this continuation, the real bounce quantities continue to the functions $X(D)$, $\omega(D)$, and $t_0(D)$ already introduced for the complex bounce. In particular, as seen from the asymptotic form in eq.\eqref{eq:t0-asymptotic}, $t_0$ acquires a specific imaginary shift for the complex bounce, so that $\sinh^2(\omega t_0)$ is minus that which appears the real bounce determinant expression. Hence, 
\begin{equation}\label{eq:det-ratio-cb-def-expanded}
\lim_{t \to \infty}\frac{\det'\hat{\mathcal{M}}}{\det\hat{\mathcal{M}}_0}
\;=\;
-
\frac{\langle \dot{z}_{\rm cl}\mid \dot{z}_{\rm cl} \rangle}{32 X^2 \omega^3 \sinh^2(\omega t_0)}
\;=\;
+\frac{D}{512}+\mathcal{O}(D^2).
\end{equation}

For the cubic potential, the single instanton is the zero-energy solution connecting the points $x=-1$ and $x=+1$. Writing the extremal equation in first-order form as $\dot{x}_I(\tau)=V'(x_I(\tau))=1-x_I(\tau)^2$, we obtain
\begin{equation}\label{eq:instanton-tanh}
x_I(\tau)=\tanh(\tau-\tau_0),\qquad \tau\in[0,t],
\end{equation}
where $\tau_0\in(0,t)$ fixes the instanton centre. The corresponding anti-instanton is obtained by time reversal and has the same determinant.

For later comparison, the normalized single-instanton determinant is
\begin{equation}\label{eq:inst-det}
\lim_{t \to \infty}\frac{1}{\langle \dot{x}_I\mid \dot{x}_I\rangle}\frac{\det'\hat{\mathcal{M}}_{\rm I}}{\det \hat{\mathcal{M}}_{0}}
= \frac{1}{64},
\qquad
\hat{\mathcal{M}}_{\rm I}
=-\frac{\drm^2}{\drm\tau^2}+4\!\left[1-\frac{3}{2}\,\sech^2(\tau-\tau_0)\right].
\end{equation}

In the exact complex bounce solution, eq.\eqref{eq:complex bounce}, $z_{\rm cl}(\tau)$ is built from the same $\tanh$ profiles as eq.\eqref{eq:instanton-tanh}, combined into an instanton-anti-instanton configuration with complex separation. In the dilute limit, this leads to the factorized form
\begin{equation}\label{eq:factorization}
\lim_{t\to\infty}\frac{1}{\langle \dot{z}_{\rm cl}\mid \dot{z}_{\rm cl}\rangle}\frac{\det' \hat{\mathcal{M}}}{\det \hat{\mathcal{M}}_0}
=
8 D\left[\lim_{t\to\infty}\frac{1}{\langle \dot{x}_I\mid \dot{x}_I\rangle}\frac{\det'\hat{\mathcal{M}}_{\rm I}}{\det \hat{\mathcal{M}}_{0}}\right]^2.
\end{equation}
The extra factor $8D$ in eq.\eqref{eq:factorization} indicates the presence of a mode of $\hat{\mathcal M}$ whose eigenvalue is small for $D\ll 1$. \edit{Following refs.\cite{behtash2015b,behtash2015,behtash2016,behtash2018},} we refer to this mode as the quasi-zero mode (QZM). It should therefore be treated by isolating the corresponding collective coordinate rather than by a naive Gaussian approximation. We therefore decompose the fluctuations as
\begin{equation}\label{eq:mode_split_qzm}
y(\tau)=a_{\rm ezm}\,y_{\rm ezm}(\tau)+a_{\rm qzm}\,y_{\rm qzm}(\tau)+y_\perp(\tau),
\qquad
y_{\rm qzm}(\tau)\propto \partial_{t_0} z_{\rm cl}(\tau;t_0),
\end{equation}
where $t_0$ is the separation parameter with fixed value given in eq.\eqref{eq:t0-exact}. In the present notation, the mode coefficient $a_1$ in eq.\eqref{eq:mode_split_ezm} is relabelled as $a_{\rm qzm}$. We now turn to a more careful treatment of this mode.
\section{Integration over the quasi-zero mode}

In the weak-noise limit $D\to 0$, the complex bounce is well approximated by a widely separated instanton-anti-instanton configuration. The action then has an almost flat direction associated with varying the instanton-anti-instanton separation, corresponding to a QZM of the fluctuation operator with a small eigenvalue, \edit{which  can be calculated from eq.\eqref{eq:fluct-operator} using standard perturbation theory techniques \cite{bender}.} To leading order in $D$, it is
\begin{equation}\label{eq:qzm_eigenvalue}
\lambda_{\rm QZM}=12D.
\end{equation}
%We isolate the corresponding collective coordinate and evaluate the resulting one-dimensional integral on the appropriate steepest-descent contour. Picard--Lefschetz theory fixes this contour \cite{witten2010,pham2011} and thereby determines the correct quasi-zero mode contribution.
%
\edit{To isolate the collective coordinate corresponding to the QZM,} 
we parameterize the separation by $\theta$, which controls the overlap of the instanton and anti-instanton tails. In the large-separation regime, the complex configuration is well approximated by an instanton-anti-instanton pair, which we denote by $Z_{\rm IA}(\tau,\theta)$:
\begin{equation}\label{eq:IA-ansatz}
Z_{\rm IA}(\tau,\theta)
=
-1-\tanh\!\left(\tau-\frac{\theta}{2}\right)
+\tanh\!\left(\tau+\frac{\theta}{2}\right).
\end{equation}
Evaluating the action on eq.\eqref{eq:IA-ansatz} gives
\begin{equation}\label{eq:qzm_action_decomp}
\mathcal S[Z_{\rm IA}]
=
2\mathcal S_I - 4Dt +\mathcal V(\theta),
\end{equation}
where $\mathcal S_I=8/3$ is the single-instanton action and $\mathcal V$ is the interaction potential. In the large-$\Re\theta$ limit,
\begin{equation}\label{eq:QZM-potential}
\mathcal V(\theta)
=
-32\,\exp(-\omega \theta)
+4D\,\omega\,\theta
+\mathcal O\!\big(\exp(-2\omega\theta)\big),
\qquad \omega \approx 2.
\end{equation}
The first term in eq.\eqref{eq:QZM-potential} is the exponentially small attraction due to tail overlap, while the linear term is generated by the It\^{o} term in the effective potential. For an instanton-instanton configuration the sign of the exponential term is reversed, corresponding to a repulsive interaction.

The critical point in the separation direction is determined by $\mathcal V'(\theta_{\rm min})=0$. For $D>0$ it lies off the real $\theta$ axis,
\begin{equation}\label{eq:qzm-saddle}
\theta_{\rm min}
=
\frac{1}{\omega}\log\!\left(\frac{8}{D}\right)
+
(2\sigma-1)\frac{i\pi}{\omega},
\qquad
\sigma\in\{0,1\},
\end{equation}
in agreement with the leading-order expansion of $t_0$ in eq.\eqref{eq:t0-asymptotic}. This is the origin of the imaginary shift of the quasi-zero mode contour\footnote{For instanton-instanton configurations, such as those arising in barrier-crossing in a double-well potential, there is no imaginary shift, and $\mathbb{R}$ is already the correct contour for the quasi-zero mode direction.}.

\begin{figure}[t]
  \centering
  \includegraphics[width=0.82\linewidth]{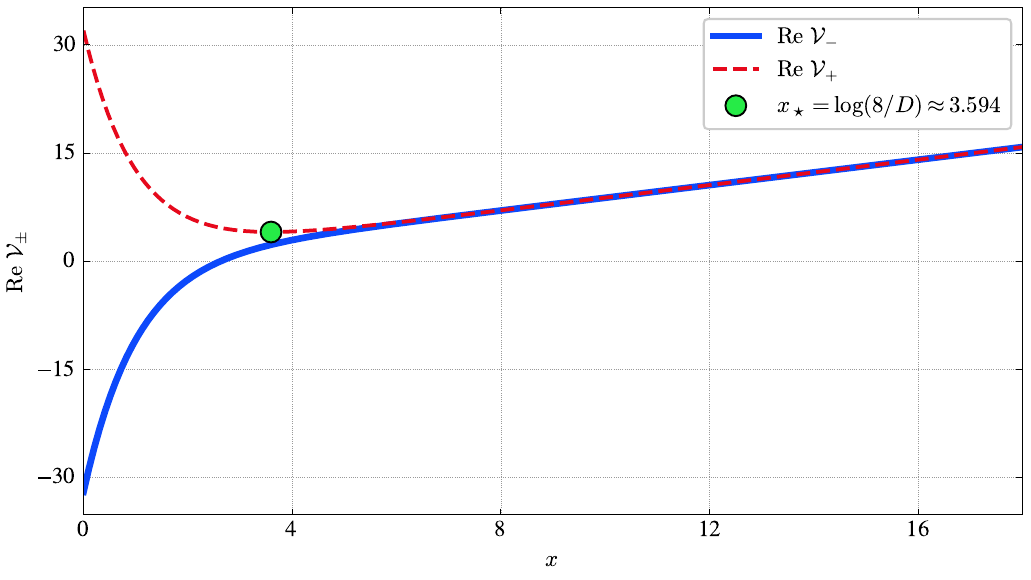}
  \caption{
    Real part of the leading-order quasi-zero mode interaction potential $\mathcal V$ plotted against the scaled separation $x:=\omega\,\Re\theta$.
    Along the real $\theta$ axis one has $\Re\mathcal V=-32\,\exp(-x)+4Dx$ (solid), which is monotone for $D>0$ and therefore has no real stationary point.
    Along the shifted contour $\theta=\Re\theta\pm i\pi/\omega$, the overlap term changes sign, giving $\Re\mathcal V=+32\,\exp(-x)+4Dx$ (dashed) with a real critical point at $x_\star=\log(8/D)$ (marked).
  }
  \label{fig:qzm-interaction-potential}
\end{figure}

Near the complex bounce solution, the integration contour factorizes into the product of the collective-coordinate directions and the hard fluctuation directions. Here ``hard'' denotes the remaining nonzero modes of $\hat{\mathcal M}$, whose eigenvalues stay well away from zero and which are treated by the usual functional Gaussian expansion. After extracting the EZM and QZM, the local measure may therefore be written schematically as
\begin{equation}\label{eq:measure-factorization}
\mathcal D z\ \sim\ \drm t_c\ \drm\theta\ \prod_{n=2}^{\infty}\drm a_n.
\end{equation}
In eq.\eqref{eq:ito-path-integral_prefactor}, where only the exact zero mode is removed, the hard modes are treated in this functional Gaussian way, while the quasi-zero mode is represented only through the quadratic approximation to the interaction potential in the separation variable $\theta$. Expanding about the critical point gives
\begin{equation}\label{eq:qzm-gauss-expansion}
\mathcal V(\theta)
=
\mathcal V(\theta_{\rm min})
+\frac{1}{2}\mathcal V''(\theta_{\rm min})(\theta-\theta_{\rm min})^2
+\cdots,
\qquad
\mathcal V'(\theta_{\rm min})=0,
\end{equation}
and hence the corresponding one-dimensional Gaussian contribution is
\begin{equation}\label{eq:QZM-integral-gauss-def}
\mathcal I_{\rm Gauss}
=
\frac{1}{\sqrt{4\pi D}}
\exp\!\left(-\frac{\mathcal V(\theta_{\rm min})}{4D}\right)
\int_{-\infty}^{\infty}\drm\xi\,
\exp\!\left(-\frac{\mathcal V''(\theta_{\rm min})}{8D}\,\xi^2\right).
\end{equation}

The exact quasi-zero mode contribution is instead defined by the contour integral
\begin{equation}\label{eq:QZM-integral-setup}
\mathcal I_{\rm QZM}
=
\frac{1}{\sqrt{4\pi D}}
\int_{\mathcal J_\sigma}\drm\theta\,
\exp\!\left(-\frac{\mathcal V(\theta)}{4D}\right),
\end{equation}
where $\mathcal J_\sigma$ is the steepest-descent contour through the relevant complex critical point.\footnote{The change of variables from the quasi-zero mode coefficient $a_{\rm qzm}$ to the separation variable $\theta$ carries a Jacobian. Since the same Jacobian appears in both the exact contour integral and its local Gaussian approximation (as well as the factor of $(4 \pi D)^{-\tfrac{1}{2}}$), it cancels in the ratio $\mathcal I_{\rm QZM}/\mathcal I_{\rm Gauss}$.}
The prefactor $(4\pi D)^{-1/2}$ appears because we are performing one fewer Gaussian integral in the numerator of eq.\eqref{eq:P_normalised_def}. Replacing the local Gaussian approximation in $\theta$ by the exact contour integral gives
\begin{multline}\label{eq:ito-path-integral_prefactor_proper}
\mathcal P(-1,t\mid -1,0)
\sim
\mathcal P_0(t)\,
\exp\!\left(+\frac{\mathcal S_0}{4D}\right)
\exp\!\left(-\frac{\mathcal S_{\rm cl}}{4D}\right)
\sqrt{\frac{\langle \dot z_{\rm cl}\mid \dot z_{\rm cl}\rangle}{4\pi D}}
\int_0^t\!\drm t_c\;
\left(\frac{\det{}'\hat{\mathcal M}}{\det \hat{\mathcal M}_0}\right)^{-1/2}
\frac{\mathcal I_{\rm QZM}}{\mathcal I_{\rm Gauss}}.
\end{multline}

To determine $\mathcal J_\sigma$, recall that for an integral of Laplace type \cite{debruijn1981} the steepest-descent contour through a critical point is characterized by $\Im\mathcal V(\theta)=\Im\mathcal V(\theta_{\rm min})$ and by $\Re\mathcal V(\theta)$ increasing away from $\theta_{\rm min}$, so that the integrand decays maximally. For the complex bounce, the steepest-descent contour through $\theta_{\rm min}$ is the shifted line
\begin{equation}\label{eq:thimble-def}
\mathcal J_\sigma
=
\mathbb R+(2\sigma-1)\,i\frac{\pi}{\omega}.
\end{equation}
To evaluate the exact integral along $\mathcal J_\sigma$, parameterize $\theta=s+(2\sigma-1)i\pi/\omega$ with $s\in\mathbb R$, so that $\exp(-\omega\theta)=-\exp(-\omega s)$ and $\drm\theta=\drm s$. Then
\begin{equation}\label{eq:qzm-gamma-step}
\int_{\mathcal J_\sigma}\!\drm\theta\,
\exp\!\left(-\frac{\mathcal V(\theta)}{4D}\right)
=
\exp\!\left[-(2\sigma-1)i\pi\right]
\int_{-\infty}^{\infty}\!\drm s\,
\exp\!\left(-\omega s-\frac{8}{D}\exp(-\omega s)\right).
\end{equation}
With the substitution $u=(8/D)\exp(-\omega s)$, eq.\eqref{eq:qzm-gamma-step} gives
\begin{equation}\label{eq:qzm-gamma-step2}
\mathcal I_{\rm QZM}
=
\frac{1}{\sqrt{4\pi D}}\,
\frac{D}{8\omega}\exp\!\left[-(2\sigma-1)i\pi\right]\Gamma(1).
\end{equation}
Since $\exp[-(2\sigma-1)i\pi]=\exp(\pm i\pi)=-1$, the phase is independent of $\sigma$.

We now evaluate the Gaussian quasi-zero mode approximation defined in eq.\eqref{eq:QZM-integral-gauss-def}. Using the leading terms in eq.\eqref{eq:QZM-potential}, one finds $\mathcal V''(\theta_{\rm min})=4D\,\omega^2$, where we used $\exp(-\omega\theta_{\rm min})=-D/8$, while $\mathcal V(\theta_{\rm min})=4D(1+\omega\theta_{\rm min})$, so that $\exp[-\mathcal V(\theta_{\rm min})/(4D)]=\exp(-1)\exp(-\omega\theta_{\rm min})$. Substituting these into eq.\eqref{eq:QZM-integral-gauss-def} gives
\begin{equation}\label{eq:thimble-integral-gaussian}
\mathcal I_{\rm Gauss}
=
\frac{1}{\sqrt{4\pi D}}\,
\frac{D}{8\omega}\exp\!\left[-(2\sigma-1)i\pi\right]\frac{\sqrt{2\pi}}{\mathrm e}.
\end{equation}
Combining this with eq.\eqref{eq:qzm-gamma-step2}, we obtain
\begin{equation}\label{eq:qzm-ratio-final}
\frac{\mathcal I_{\rm QZM}}{\mathcal I_{\rm Gauss}}
=
\frac{\mathrm e}{\sqrt{2\pi}}.
\end{equation}
The spurious factor $\sqrt{2\pi}/\mathrm e$ is seen as the leading-order Stirling approximation to $\Gamma(1)$ -- \edit{the Gaussian approximation.}
%Picard--Lefschetz theory removes this error by selecting the correct quasi-zero mode contour and thereby restores the exact value $\Gamma(1)=1$.

% Recovering the escape rate
% ------------------------------------------------------------
% ------------------------------------------------------------
% Extracting the escape rate
% ------------------------------------------------------------
The escape rate can now be extracted directly, with no additional $D\mapsto -D$ continuation. In the dilute limit, the coarse-grained survival probability $P_{\rm well}$ has the form
\begin{equation}\label{eq:linear_rate_def}
P_{\rm well}(t)=1-\Gamma_K t+\mathcal O\!\big((\Gamma_K t)^2\big).
\end{equation}
To identify $\Gamma_K$, we coarse-grain near the stable minimum at $x=-1$ (\cite{mckane3}). In the weak-noise limit, particles that remain near $x=-1$ are confined to a capture region of width $\ell_{\rm well}\sim \sqrt{D/V''(-1)}$, where the potential is effectively harmonic. Once particles in this region have quasi-equilibrated, the density is of the form
\begin{equation}\label{eq:local_factorization}
\mathcal P(x,t\mid -1,0)\simeq P_{\rm well}(t)\,p_{\rm eq}(x),
\qquad
p_{\rm eq}(x)=\sqrt{\frac{V''(-1)}{2\pi D}}\,
\exp\!\left(-\frac{V''(-1)}{2D}(x+1)^2\right),
\end{equation}
with $p_{\rm eq}$ normalized. Evaluating eq.\eqref{eq:local_factorization} at $x=-1$ gives
\begin{equation}\label{eq:cg_prob_from_point}
P_{\rm well}(t)\simeq \mathcal P(-1,t\mid -1,0)\,\sqrt{\frac{2\pi D}{V''(-1)}}.
\end{equation}
In the one-event term of eq.\eqref{eq:ito-path-integral_prefactor_proper}, the translation integral $\int_0^t \drm t_c=t$ produces the factor of $t$. Comparing eqs.\eqref{eq:linear_rate_def}, \eqref{eq:cg_prob_from_point}, and \eqref{eq:ito-path-integral_prefactor_proper}, we obtain
\begin{equation}\label{eq:rate-cubic-explicit}
-\Gamma_K(D)t=
\left(\frac{\mathrm e}{\sqrt{2\pi}}\right)
\sqrt{\frac{8\omega^3}{\pi D}}\,
X(D)\,
\exp\!\left(-\frac{Y(D)}{4D}\right)\,
\bigl(\sqrt{Z(D)+1}-\sqrt{Z(D)}\bigr)^4\,
\sqrt{1-4Q(D)^2}\,t,
\end{equation}
where the factor $\mathrm e/\sqrt{2\pi}$ is precisely the ratio $\mathcal I_{\rm QZM}/\mathcal I_{\rm Gauss}$ obtained above. The right-hand side of eq.\eqref{eq:rate-cubic-explicit} is negative, so $\Gamma_K(D)$ is positive.

As $D\to 0$, the quantities entering eq.\eqref{eq:rate-cubic-explicit} satisfy
\begin{equation}\label{eq:rate-building-blocks-asymptotic}
Y\to \frac{16}{3},
\qquad
\bigl(\sqrt{Z+1}-\sqrt{Z}\bigr)^4\to -\,\frac{D}{8},
\qquad
\sqrt{1-4Q^2}\to \sqrt{\frac{2}{D}},
\qquad
X\to 1,
\qquad
\omega\to 2.
\end{equation}
Substituting these limits into eq.\eqref{eq:rate-cubic-explicit} gives
\begin{equation}\label{eq:rate-asymptotic}
\Gamma_K(D)=\frac{1}{\pi}\exp\!\left(-\frac{4}{3D}+\mathcal O(D)\right)\Bigl(1+\mathcal O(D)\Bigr),
\end{equation}
whose leading term agrees with the Kramers formula
\begin{equation}\label{eq:kramers}
\Gamma_K=\frac{\sqrt{|V''(x_{\min})V''(x_{\max})|}}{2\pi}\,
\exp\!\left(-\frac{\Delta V}{D}\right),
\end{equation}
since $\Delta V=4/3$ and $\sqrt{|V''(-1)V''(+1)|}=2$. Higher-order corrections in $D$ appear in both the exponent and the prefactor because the It\^{o} action itself contains terms of order $D$. We stress that the cubic potential has been chosen for clarity. The explicit rate formula in eq.\eqref{eq:rate-cubic-explicit} is model-dependent, but the quasi-zero mode analysis in the It\^{o} formulation extends more generally. For attractive interaction potentials, the ratio of the exact integral to its local Gaussian approximation is given by the general factor in~\ref{App:generalQZM}. For the cubic model, $ B=1$, so this reduces to the factor $\mathrm e/\sqrt{2\pi}$ in eq.\eqref{eq:rate-cubic-explicit}.
\begin{figure}
\centering
\includegraphics[width=0.8\textwidth]{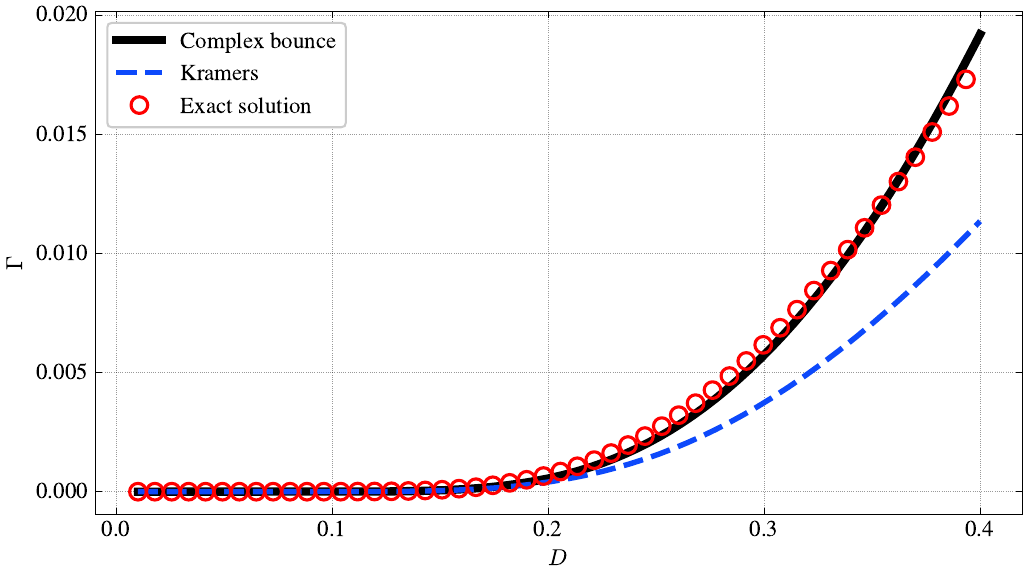}
\caption{Barrier-crossing rate $\Gamma$ against noise strength $D$. Red circles show the exact rate calculated with eq.\eqref{eq:MFPT}, the dashed blue curve shows the Kramers approximation eq.\eqref{eq:kramers}, and the solid black curve is the complex bounce rate eq.\eqref{eq:rate-cubic-explicit}.}
\label{fig:rate1}
\end{figure}
Fig.\ref{fig:rate1} shows the complex bounce rate eq.\eqref{eq:rate-cubic-explicit}, the Kramers rate eq.\eqref{eq:kramers} and the exact rate, determined by evaluating the integral formula below numerically, against the noise strength $D$. 
\eq
\Gamma_{\rm exact} = D\left(\int_{-1}^1 \exp\left(\frac{V(x')}{D}\right) \int_{-\infty}^{x'} \exp\left(\frac{-V(x)}{D}\right)\dee x\,\dee x'\right)^{-1}.\label{eq:MFPT}
\eeq The complex bounce rate formula agrees with the exact rate somewhat better than the standard Kramers approximation up to moderate values of $D$.

\section{Conclusion}
We have shown that, in order to obtain the correct steepest-descent contour for the quasi-zero mode integral, one must work in the It\^{o} formulation of the stochastic path integral. The reason is that the necessary $D$-dependence enters directly through the It\^{o} variational equations. In this formulation, complex extremal solutions exist, and for the cubic model we have derived them explicitly. Working with these solutions, we resolved the problem of analytic continuation in the quasi-zero mode integral and derived the Kramers rate. Although the cubic model was chosen for clarity, the mechanism itself is not restricted to this example and should extend to more general potentials.

The main quantitative result is the explicit rate formula in eq.\eqref{eq:rate-cubic-explicit}. More broadly, for the present class of additive white-noise problems, the analysis shows that the combination of It\^{o} calculus and complex extremal solutions provide a coherent framework for the weak-noise expansion, \edit{provided the appropriate steepest-descent contours are used. In general, these contours are determined by Picard--Lefschetz theory \cite{witten2010,pham2011}.
}
It would be interesting to extend this analysis to more general potentials and, with appropriate care, to higher-dimensional systems with additional soft directions. It would also be useful to understand how far the present mechanism persists in more general nonequilibrium settings, and whether an analogous geometric picture survives for coloured-noise problems.

\subsection*{Acknowledgments}
SPF acknowledges many useful discussions with Profs A. Archer and T. Ala-Nissila. This work was supported by the UK EPSRC, grant numbers EP/R005974/1 and UKRI3643, and the Leeds Doctoral Scholarship.

\end{spacing}

%\bibliography{paper}

\providecommand{\newblock}{}

\begin{spacing}{1.25}

\appendix
\section{The complex bounce solution and its action}\label{App:CBmaterial}
The integral in eq.\eqref{eq:cb-quadrature} can be performed by a sequence of elementary transformations. These were computed in Refs. \cite{behtash2015b}, and are reproduced here for completeness as there is a slight difference resulting from differing definitions of the mass of the effective Hamiltonian particle.  
% Introduce the translated variables
% \eq
% y:=z-z_{\rm cr},
% \qquad
% a:=z_{\rm tu}^{+}-z_{\rm cr},
% \qquad
% b:=z_{\rm tu}^{-}-z_{\rm cr},
% \eeq
% so that $\Phi(z)=(z-z_{\rm cr})^{2}(z-z_{\rm tu}^{+})(z-z_{\rm tu}^{-})$ implies
% \eq\label{eq:y-ode}
% \dot y^{\,2}=y^{2}(y-a)(y-b).
% \eeq
% Applying the reciprocal substitution $w:=1/y$ reduces \eqref{eq:y-ode} to
% \eq\label{eq:w-ode}
% \dot w^{\,2}=ab\,w^{2}-(a+b)w+1.
% \eeq
% Taking $w=w_{0}+v$ with $w_{0}:=\frac{a+b}{2ab}$ gives
% \eq\label{eq:v-ode}
% \dot v^{\,2}=ab\,v^{2}+c,
% \qquad
% c:=-\frac{(a-b)^{2}}{4ab}.
% \eeq
% Defining the frequency by 
% \eq\label{eq:omega-def}
% \omega^{2}:=ab=(z_{\rm tu}^{+}-z_{\rm cr})(z_{\rm tu}^{-}-z_{\rm cr}),
% \eeq
% equation \eqref{eq:v-ode} integrates to
% \eq\label{eq:v-sol}
% v(\tau)=\frac{\sqrt{c}}{\omega}\,\sinh\!\big(\omega(\tau-\tau_{*})\big),
% \eeq
% for a (generally complex) constant $\tau_{*}$. Inverting the substitutions $v\mapsto w\mapsto y\mapsto z$
One finds
% \eq\label{eq:z-sinh-form}
% z_{\rm cl}(\tau)
% =
% z_{\rm cr}
% +
% \frac{1}{w_{0}+\displaystyle \frac{\sqrt{c}}{\omega}\,\sinh\!\big(\omega(\tau-\tau_{*})\big)}.
% \eeq
\eq\label{eq:z-rational-cosh}
z_{\rm cl}(\tau)
=
z_{\rm cr}
+
(z_{\rm tu}^{\pm}-z_{\rm cr})\,
\frac{1+C}{\cosh(\omega\tau)+C}
\eeq
where 
\eq\label{eq:C-def}
C:= -\frac{(z_{\rm tu}^{+}-z_{\rm cr})+(z_{\rm tu}^{-}-z_{\rm cr})}{(z_{\rm tu}^{+}-z_{\rm cr})-(z_{\rm tu}^{-}-z_{\rm cr})} = \cosh(\omega t_{0}),
\eeq
defining $t_0$. Note that $C$ is purely imaginary when $z_{\rm tu}^{-}=\overline{z_{\rm tu}^{+}}$.
%
% where we have imposed the turning condition $z_{\rm cl}(0)=z_{\rm tu}^{\pm}$ fixes $\tau_{*}$ and is encoded by
%
% which  With this choice one obtains the compact cosh--rational form
%
% Introducing $t_{0}$ via
% \eq\label{eq:t0-from-C}
% \cosh(\omega t_{0})=C,
% \qquad
% t_{0}=\frac{1}{\omega}\,\arcosh(C),
% \eeq
Using the identity
\eq\label{eq:tanh-identity}
\frac{1+C}{\cosh(\omega\tau)+C}
=
\frac{1}{2}\coth\!\Big(\frac{\omega t_{0}}{2}\Big)
\left[
\tanh\!\Big(\frac{\omega}{2}(\tau+t_{0})\Big)
-
\tanh\!\Big(\frac{\omega}{2}(\tau-t_{0})\Big)
\right],
\eeq
substitution into eq.\eqref{eq:z-rational-cosh} gives
\eq
z_{\rm cl}(\tau)
=
z_{\rm cr}
+
X(D)\left[
\tanh\!\Big(\frac{\omega}{2}(\tau+t_{0})\Big)
-
\tanh\!\Big(\frac{\omega}{2}(\tau-t_{0})\Big)
\right],
\eeq
with
\eq\label{eq:X-from-turning}
X(D)
=
\frac{z_{\rm tu}^{\pm}-z_{\rm cr}}{2}\,
\coth\!\Big(\frac{\omega t_{0}}{2}\Big),
\eeq
which matches the form stated in eq.\eqref{eq:complex bounce}. To derive the complex bounce action, note that along an extremal, one has the first integral
\begin{equation}\label{eq:cb_quadrature_app}
\dot z^{\,2}=\Phi(z),
\qquad
\Phi(z):=H+V'(z)^{2}-2D\,V''(z)=H-U(z),
\end{equation}
so that
\begin{equation}\label{eq:cb_action_quadrature_app}
\mathcal{S}_{\rm cl}
=
-Ht+2\int_{\gamma_{\rm cb}}\sqrt{\Phi(z)}\,\drm z,
\end{equation}
using $\drm\tau=\drm z/\dot z=\drm z/\sqrt{\Phi(z)}$ and where
\begin{equation}\label{eq:Phi_cubic_app}
\Phi(z)=H+(a^{2}-z^{2})^{2}+4Dz = (z-z_{\rm cr})^{2}(z-z_{\rm tu}^{-})(z-z_{\rm tu}^{+}) .
\end{equation}
The integration contour,$\gamma_{\rm cb}$, denotes one full complex bounce traversal, i.e. the out-and-back path $z_{\rm cr}\to z_{\rm tu^{\pm}}\to z_{\rm cr}$, so that
$\displaystyle \int_{\gamma_{\rm cb}}\sqrt{\Phi(z)}\,\drm z
=2\int_{z_{\rm cr}}^{z_{\rm tu^{\pm}}}\sqrt{\Phi(z)}\,\drm z$. Hence, accounting for this factor of 2, eq.\eqref{eq:cb_action_quadrature_app} is elementary and can be integrated directly, giving
\begin{equation}\label{eq:cb_action_exact_alt_app}
\mathcal{S}_{\rm cl}
=
-Ht
+\frac{16a^{3}}{3}\sqrt{1-\frac{3D}{2a^{2}z_{\rm cr}}}
+16D\log\!\left(
\sqrt{\frac{1}{2}+\sqrt{\frac{z_{\rm cr}^{3}}{2D}}}
+
\sqrt{-\frac{1}{2}+\sqrt{\frac{z_{\rm cr}^{3}}{2D}}}
\right).
\end{equation}
It is convenient to parameterize the remaining quantities by
\begin{align}
Q(D)
&:= \frac{z_{\rm cr}}{\sqrt{2-2z_{\rm cr}^{2}}},
\\[0.25em]
Y(D)
&:= \frac{16}{3}\sqrt{1-\frac{3D}{2z_{\rm cr}}},
\\[0.25em]
Z(D)
&:= -\frac{1}{2}+\sqrt{\frac{z_{\rm cr}^{3}}{2D}},
\\[0.25em]
\omega(D)
&:= \sqrt{6z_{\rm cr}^{2}-2},
\end{align}
and then our solutions match the extremal solution and action given in the main text in eq.\eqref{eq:complex bounce} and eq.\eqref{eq:complex-action}.

\section{The complex bounce fluctuation operator}\label{App:FDmaterial}
\subsection*{Derivation of the fluctuation operator eq.\eqref{eq:fluct-operator}.}
\label{App:fluct_operator_derivation}
Along a classical solution $z_{\rm cl}$ the Euler--Lagrange equation takes the form
\begin{equation}
\label{eq:EL_newton_app}
\ddot z_{\rm cl}(\tau)=-\frac12\,U'(z_{\rm cl}(\tau)),
\end{equation}
and the fluctuation operator is
\begin{equation}
\label{eq:M_general_app}
\hat{\mathcal{M}}
=
-\frac{\drm^2}{\drm\tau^2}-\frac12\,U''(z_{\rm cl}(\tau)).
\end{equation}
Differentiating eq.\eqref{eq:EL_newton_app} with respect to $\tau$ gives
\begin{equation}
\label{eq:ratio_trick_app}
\dddot z_{\rm cl}(\tau)
=
-\frac12\,U''(z_{\rm cl}(\tau))\,\dot z_{\rm cl}(\tau),
\end{equation}
hence (away from isolated zeros of $\dot z_{\rm cl}$, with extension by continuity)
\begin{equation}
\label{eq:ratio_identity_app}
-\frac12\,U''(z_{\rm cl}(\tau))
=
\frac{\dddot z_{\rm cl}(\tau)}{\dot z_{\rm cl}(\tau)}.
\end{equation}
For the complex bounce eq.\eqref{eq:complex bounce},
\begin{equation}
\label{eq:zcl_tanhpair_app}
z_{\rm cl}(\tau)
=
z_{\rm cr}
+
X\Big[
\tanh\!\Big(\frac{\omega}{2}(\tau+t_0)\Big)
-
\tanh\!\Big(\frac{\omega}{2}(\tau-t_0)\Big)
\Big],
\end{equation}
applying eq.\eqref{eq:ratio_identity_app} to eq.\eqref{eq:zcl_tanhpair_app} and simplifying gives
\begin{equation}
-\frac12\,U''(z_{\rm cl}(\tau))
=
\omega^2\left[
1-\frac32\left(
\sech^2\!\Big(\frac{\omega}{2}(\tau+t_0)\Big)
+
\sech^2\!\Big(\frac{\omega}{2}(\tau-t_0)\Big)
\right)\right],
\end{equation}
and substitution into eq.\eqref{eq:M_general_app} yields eq.\eqref{eq:fluct-operator}.

\section{Determinant formula for the real bounce}\label{App:RBdet}

In this appendix we summarize the determinant formula used for the real bounce in eq.\eqref{eq:det-ratio-rb-closed}. We first state the general result for a Hermitian one-dimensional fluctuation operator \cite{tarlie}, and then specialize it to the real bounce used in the main text.

Let $x_{\rm cl}(\tau)$ be a real bounce on a large interval $\tau\in[-t/2,t/2]$, returning to the same fixed point as $\tau\to\pm\infty$, and consider the Hermitian fluctuation operator
\begin{equation}\label{eq:App-M-herm}
\hat{\mathcal M}
=
-\,\frac{\drm^2}{\drm\tau^2}
-\frac{1}{2}\,U''\!\bigl(x_{\rm cl}(\tau)\bigr),
\end{equation}
with
\begin{equation}\label{eq:App-U-def}
U(x)=-V'(x)^2+2D\,V''(x).
\end{equation}
To normalize the determinant, we introduce the harmonic operator
\begin{equation}\label{eq:App-M0-def}
\hat{\mathcal M}_0
=
-\,\frac{\drm^2}{\drm\tau^2}
+\omega_0^2,
\qquad
\omega_0^2=-\frac{1}{2}\,U''(x_{\rm cr}),
\end{equation}
where $x_{\rm cr}$ is the fixed point approached at both endpoints. The one-dimensional determinant formula with the exact zero mode removed then gives
\begin{equation}\label{eq:App-master-finite}
\frac{\det\nolimits'\hat{\mathcal M}}{\det \hat{\mathcal M}_0}
=
-\,\frac{\omega_0}{4\,\sinh(\omega_0 t)}\,
\frac{\langle \dot{x}_{\rm cl}\mid \dot{x}_{\rm cl}\rangle}
{\ddot{x}_{\rm cl}(-t/2)\,\ddot{x}_{\rm cl}(t/2)}.
\end{equation}
For a real bounce, the exact zero mode is $\dot{x}_{\rm cl}$, and its asymptotic form is
\begin{equation}\label{eq:App-rb-asymp}
\dot{x}_{\rm cl}(\tau)\sim
\alpha_-\,\exp(\omega_0\tau)
\quad (\tau\to-\infty),
\qquad
\dot{x}_{\rm cl}(\tau)\sim
-\alpha_+\,\exp(-\omega_0\tau)
\quad (\tau\to+\infty),
\end{equation}
with $\alpha_\pm>0$. Hence
\begin{equation}\label{eq:App-ddot-asymp}
\ddot{x}_{\rm cl}(-t/2)\sim
\omega_0\alpha_-\,\exp(-\omega_0 t/2),
\qquad
\ddot{x}_{\rm cl}(t/2)\sim
\omega_0\alpha_+\,\exp(-\omega_0 t/2),
\end{equation}
while
\begin{equation}\label{eq:App-sinh-asymp}
\sinh(\omega_0 t)\sim \frac{1}{2}\,\exp(\omega_0 t)
\qquad (t\to\infty).
\end{equation}
Substituting these asymptotics into eq.\eqref{eq:App-master-finite} yields
\begin{equation}\label{eq:App-master-infinite}
\lim_{t\to\infty}
\frac{\det\nolimits'\hat{\mathcal M}}{\det \hat{\mathcal M}_0}
=
-\frac{1}{2\,\omega_0}\,
\frac{\langle \dot{x}_{\rm cl}\mid \dot{x}_{\rm cl}\rangle}
{\alpha_-\alpha_+}.
\end{equation}

We now specialize to the symmetric real bounce used in the main text. Its exact form is
\begin{equation}\label{eq:rb-exact-form}
x_{\rm rb}(\tau)
=
x_{\rm cr}
+
X_{\rm rb}(D)\!\left[
\tanh\!\left(\frac{\omega_{\rm rb}}{2}(\tau+t_{0,\rm rb})\right)
-
\tanh\!\left(\frac{\omega_{\rm rb}}{2}(\tau-t_{0,\rm rb})\right)
\right],
\end{equation}
where
\begin{equation}\label{eq:App-Xrb-def}
X_{\rm rb}(D)
=
\frac{x_{\rm tu}-x_{\rm cr}}{2}\,
\coth\!\left(\frac{\omega_{\rm rb} t_{0,\rm rb}}{2}\right).
\end{equation}
The critical point is
\begin{equation}\label{eq:r-critical}
x_{\rm cr}
=
-\frac{2}{\sqrt{3}}
\cos\!\left(
\frac{1}{3}\arccos\!\left(\frac{3\sqrt{3}D}{2}\right)
-\frac{4\pi}{3}
\right)
=
1-\frac{D}{2}-\frac{3D^2}{8}+\cdots,
\end{equation}
the turning point is
\begin{equation}\label{eq:App-xtu-def}
x_{\rm tu}
=
-\,x_{\rm cr}+\sqrt{\frac{2D}{x_{\rm cr}}},
\end{equation}
and
\begin{equation}\label{eq:App-omegarb-def}
\omega_{\rm rb}
=
\sqrt{6x_{\rm cr}^2-2}.
\end{equation}
The separation is purely real,
\begin{equation}\label{eq:t0rb-exact}
t_{0,\rm rb}
=
\frac{2}{\omega_{\rm rb}}
\operatorname{arcosh}\!\left(
\sqrt{\frac{x_{\rm cr}}{\sqrt{2-2x_{\rm cr}^2}}+\frac{1}{2}}
\right).
\end{equation}
Differentiating eq.\eqref{eq:rb-exact-form} gives
\begin{equation}\label{eq:App-rb-dot}
\dot{x}_{\rm rb}(\tau)
=
\frac{\omega_{\rm rb}X_{\rm rb}}{2}
\left[
\sech^2\!\left(\frac{\omega_{\rm rb}}{2}(\tau+t_{0,\rm rb})\right)
-
\sech^2\!\left(\frac{\omega_{\rm rb}}{2}(\tau-t_{0,\rm rb})\right)
\right].
\end{equation}
From this expression one finds
\begin{equation}\label{eq:App-rb-alpha}
\alpha_-=\alpha_+
=
4X_{\rm rb}\,\omega_{\rm rb}\,\sinh(\omega_{\rm rb}t_{0,\rm rb}).
\end{equation}
Since $\omega_0=\omega_{\rm rb}$ for the real bounce, eq.\eqref{eq:App-master-infinite} becomes
\begin{equation}\label{eq:App-rb-final}
\lim_{t\to\infty}
\frac{\det\nolimits'\hat{\mathcal M}_{\rm rb}}{\det \hat{\mathcal M}_0}
=
-
\frac{\langle \dot{x}_{\rm rb}\mid \dot{x}_{\rm rb}\rangle}
{32X_{\rm rb}^2\omega_{\rm rb}^3\sinh^2(\omega_{\rm rb}t_{0,\rm rb})},
\end{equation}
which is the formula quoted in eq.\eqref{eq:det-ratio-rb-closed}.

\section{General quasi-zero mode correction factor}\label{App:generalQZM}

The cubic model is useful because all quantities can be written explicitly, but consider a more general interaction potential of the form
\begin{equation}\label{eq:general-int-V}
\mathcal V_{\rm int}(\theta)
=
- A  \exp(-\omega\theta)
+4B D\,\omega\,\theta,
\qquad
A, B>0,
\end{equation}
which captures the large-separation structure of an attractive exponential overlap together with a linear It\^{o} contribution.

The critical point satisfies $\mathcal V_{\rm int}'(\theta_{\rm min})=0$, so that
\begin{equation}\label{eq:general-theta-min}
\theta_{\rm min}
=
\frac{1}{\omega}\log\!\left(\frac{A}{4 B D}\right)
+
(2\sigma-1)\frac{i\pi}{\omega},
\qquad
\sigma\in\{0,1\}.
\end{equation}
Along the corresponding steepest-descent contour
\begin{equation}
\mathcal J_\sigma=\mathbb R+(2\sigma-1)\frac{i\pi}{\omega},
\end{equation}
the exact quasi-zero mode integral is
\begin{equation}\label{eq:general-Iqzm}
\mathcal I_{\rm QZM}
=
\frac{1}{\sqrt{4\pi D}}
\int_{\mathcal J_\sigma}\drm\theta\,
\exp\!\left(-\frac{\mathcal V_{\rm int}(\theta)}{4D}\right).
\end{equation}
Parameterizing $\theta=s+(2\sigma-1)i\pi/\omega$ with $s\in\mathbb R$ and setting
\begin{equation}
u=\frac{A}{4D}\exp(-\omega s),
\end{equation}
one finds
\begin{equation}\label{eq:general-Iqzm-evaluated}
\mathcal I_{\rm QZM}
=
\frac{1}{\omega\sqrt{4\pi D}}\,
\exp\!\left[-(2\sigma-1)i\pi B\right]
\left(\frac{4D}{A}\right)^{ B}
\Gamma( B).
\end{equation}
The corresponding local Gaussian approximation is obtained by expanding about $\theta_{\rm min}$, eq.\eqref{eq:qzm-gauss-expansion}, 
% \begin{equation}
% \mathcal V_{\rm int}(\theta)
% =
% \mathcal V_{\rm int}(\theta_{\rm min})
% +\frac{1}{2}\mathcal V_{\rm int}''(\theta_{\rm min})(\theta-\theta_{\rm min})^2
% +\cdots,
% \end{equation}
with $\mathcal V_{\rm int}''(\theta_{\rm min})=4 B D\,\omega^2$ and $\mathcal V_{\rm int}(\theta_{\rm min})=4 B D\bigl(1+\omega\theta_{\rm min}\bigr)$.
Hence
\begin{equation}\label{eq:general-Igauss}
\mathcal I_{\rm Gauss}
=
\frac{1}{\omega\sqrt{4\pi D}}\,
\exp\!\left[-(2\sigma-1)i\pi B\right]
\left(\frac{4 B D}{A\mathrm e}\right)^{ B}
\sqrt{\frac{2\pi}{ B}}.
\end{equation}
Taking the ratio, the common phase and normalization factors cancel, giving
\begin{equation}\label{eq:general-error-factor}
\frac{\mathcal I_{\rm QZM}}{\mathcal I_{\rm Gauss}}
=
\frac{\Gamma( B)}
{\left(\dfrac{ B}{\mathrm e}\right)^{ B}\sqrt{2\pi/ B}}.
\end{equation}
Thus the quasi-zero mode correction factor is precisely the factor by which the exact Gamma function differs from its Stirling approximation
\begin{equation}
\Gamma(z)\sim \sqrt{\frac{2\pi}{z}}\left(\frac{z}{\mathrm e}\right)^z
\qquad (z\to\infty).
\end{equation}
For the cubic potential one has $ B=1$, and eq.\eqref{eq:general-error-factor} reduces to
\begin{equation}
\frac{\mathcal I_{\rm QZM}}{\mathcal I_{\rm Gauss}}
=
\frac{\mathrm e}{\sqrt{2\pi}},
\end{equation}
in agreement with eq.\eqref{eq:qzm-ratio-final}.

\end{spacing}

\end{document}